# Inventions on GUI for Touch Sensitive Screens
## A TRIZ based analysis

**Umakant Mishra**

http://umakantm.blogspot.in

**Contents**

# 1. Introduction

A touch sensitive screen displays the information on the screen and also receives the input by sensing a user's touch on the same screen. This mechanism facilitates system interaction directly through the screen without needing a mouse or keyboard. This method has the advantage to make the system compact by removing keyboard, mouse and similar interactive device.

Although the larger displays (640x480 or higher resolution) are generally comfortable for touch screen displays there are circumstances where a touch screen implementation becomes difficult,

- The display screens of portable devices are becoming smaller thereby leaving lesser space for display of data, menu or touch screen interaction.

- Some screens need to display so much of information that they hardly can afford any space to display touch screen buttons. In other words, the touch screen interface permanently occupies screen space thereby reducing the screen space for displaying user information.

## 2. Inventions on GUI for touch sensitive screens

### 2.1 Split screen keyboard emulator (Patent 5031119)

**Background problem**

A keyboard is large in size and reduces the portability of a computer. Two different methods are followed to solve this problem. One is to attach a small keyboard as an integral part of the computer as in a laptop. The other is to emulate a keyboard on the screen and operate with an interactive pointing device.

The second option, although good, has a practical difficulty. If the program requires many keystrokes, the user faces difficulty by frequently switching between program screen and keyboard screen.

**Solution provided by the invention**

Patent 5031119 (invented by Dulaney et al., assigned by Tandy Corporation, issued July 1991) disclosed a method of providing keystroke data to an application without utilizing a keyboard.

According to the invention, the screen display is divided into two segments. The graphics display of the application program is displayed in first segment, and the keyboard is displayed on the second segment. Keys are selected by touching the screen at the location of the graphic representation of the key on the screen.

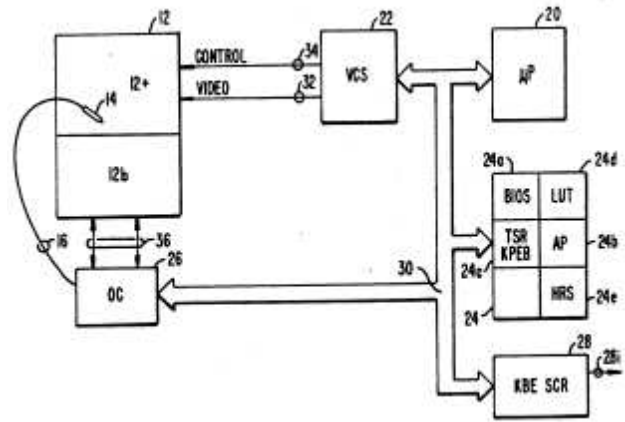

This system is efficient and transparent to the application program. Since the application program and the keyboard representation are simultaneously displayed there is no need to switch programs to provide keystroke data.

**TRIZ based analysis**

The invention substitutes a physical keyboard with a soft-keyboard displayed on the screen (Principle-26: Copying, Principle-28: Mechanics substitution).

The screen is split into two segments to display application program in one half and Keyboard on the other half (Principle-1: Segmentation).

When the keystrokes are required, the application and the keyboard are displayed simultaneously, so there is no prior-art problem of switching between application screen and keyboard screen (Principle-5: Merging).

## 2.2 Lateral alignment of Pull down menus on a LCD display using a touch sensor (Patent 5559944)

### Background

There are some disadvantages while implementing conventional menus in a touch sensitive screen. For example when the user selects the menu items through a pen or finger, the items in the pull down menu are hidden by the pen and the finger.

### Solution provided by the invention

Makoto Ono invented a method of aligning pull down menus (Patent: 5559944, Assigned to IBM, Sep 96). According to the invention the menu bar is vertically lined up at the right or left of a screen. The drop down menu items are laterally arranged in the shape of an arc for better movement of the wrist during selection.

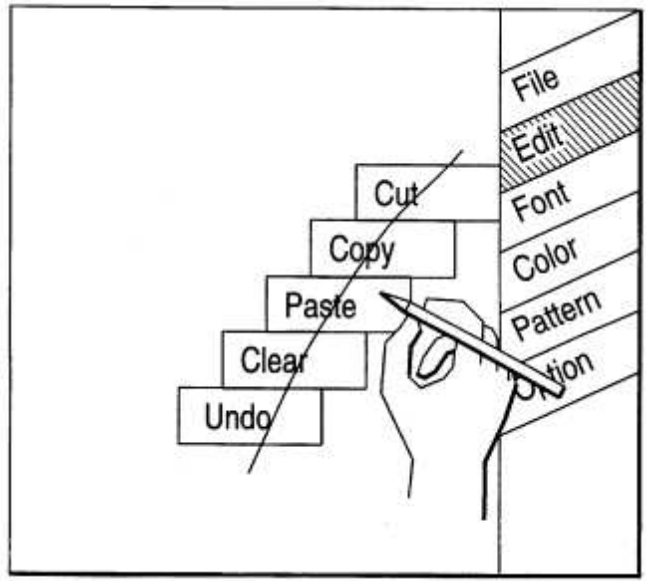

### TRIZ based analysis

The menu items should be well visible during selection (Ideal Final Result).
The pen or finger used to select the menu items should not block the visibility of the menu items (desired result).

The invention displays the menu items in a lateral direction (instead of vertical) for clear visibility even while operating with finger or pen (Principle-14: Curvature, Principle-17: Another Dimension).

## 2.3 Graphical user interface touch screen with an auto zoom feature (Patent 6211856)

**Background problem**

Today's home entertainment systems have a large number of functions available to the user. Many of them are remote controlled by GUI displays. Although a GUI adds user friendliness to these devices, typically these devices (remote controls, photocopier control panel etc.) are relatively small to display the icons in a size that can be accessible by a person's finger.

**Solution provided by the invention**

Patent 6211856 (invented by Choi et al., issued Apr 2001) used a zooming feature to solve this problem. According to the invention an entire collection of icons are displayed at a scale in which the individual function of each icon is recognizable but too small to easily access features of the function. When the user touches the icon on the screen, that icon is zoomed to the full screen so that the user can select the desired feature.

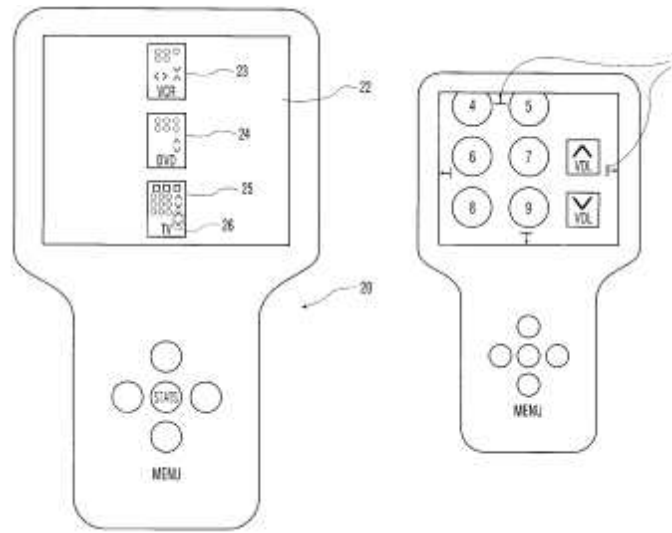

For example the initial screen may display the icons of a TV, VCR, DVD, Keyboard etc. When the user touch selects TV then the TV icon is zoomed to display the functions available inside TV, such as channels, volume etc.

**TRIZ based analysis**

The icons in these tiny touch sensitive devices should be large enough to be selected by a finger or a stylus. But there is not enough space in the screen to display them in larger size. (Contradiction).

The invention displays icons in a size that are easy to recognize but difficult to select the underlying features (Principle-16: Partial or excessive action).

According to invention, the icon is zoomed when selected so that the underlying features become selectable. (Principle-37: Expansion).

## 2.4 Method and system for providing touch sensitive screens for the visually impaired (6489951 and 6496182)

**Background problem**

A touch sensitive screen is operated by user's touch on the touch sensitive screen. This mechanism is helpful for users having full or limited eyesight but does not benefit the users who have lost their sight completely. There is a need to make the touch sensitive screen useful for the blind users.

**Solution provided by the invention**

Wong et al. disclosed a touch sensitive screen (US Patent 6489951 and 6496182, Assigned to Microsoft Corporation, Dec 2002) which is helpful for users having very limited sights or no sights. When the user touches an object, the system announces the text associated with that area through audio output. A user may execute a control, such as a button by dragging a finger onto the control and then lifting the finger out of the touch sensitive screen.

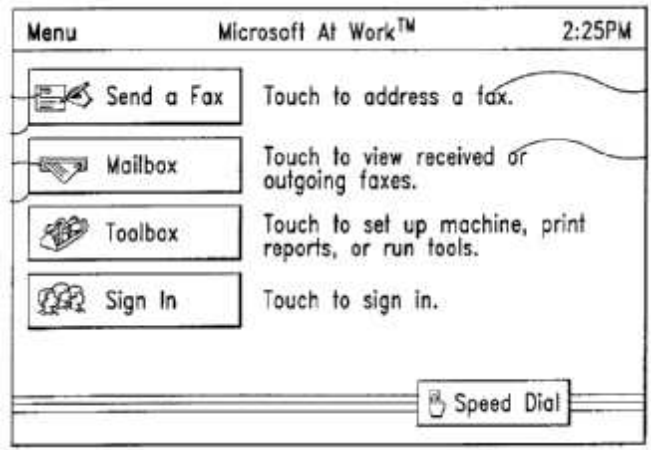

The touch sensitive screen has two modes, the scroll mode and explore mode. In scroll mode, the user can use a scroll thumb to scroll through a list, in explore mode the user can drag the finger on to a control and lift the finger of the touch sensitive screen.

**TRIZ based analysis**

The touch-sensitive screen announces objects (such as buttons) in audible voice when the user touches them (Principle-23: Feedback).

The invention uses different touching methods (such as, dragging a finger, lifting a finger after dragging etc.) to enter into different modes and doing different operations (Principle-28: Mechanics substitution).

## 3. Summary

The implementation of GUI in a touch screen interface faces various difficulties of screen space, organizing and managing touch sensitive icons on the screen and so on. The problems can be successfully eliminated by applying appropriate Inventive Principles.

For example, the problem of space in the display area may be solved by eliminating unwanted items from the screen (Principle-2: Taking out). The same can also be manage by displaying touch-screen interface on demand and not permanently (Principle-15: Dynamize). Similarly, the touch-sensitive buttons on the screen can be accompanied with audible voice to facilitate visually impaired users (Principle-23: Feedback).

Touch sensitive interface is a growing technology and GUI in the touch screen interface is a growing concern. We can expect to see more and more inventions in this field in future.